# Circular Polarization of Periodic Leaky-Wave Antennas With Axial Asymmetry: Theoretical Proof and Experimental Demonstration

Simon Otto, *Member, IEEE*, Zhichao Chen, Amar Al-Bassam, Andreas Rennings, *Member, IEEE*, Klaus Solbach, and Christophe Caloz, *Fellow, IEEE*

*Abstract*—This paper includes two contributions. First, it proves that the series and shunt radiation components, corresponding to longitudinal and transversal electric fields, respectively, are always in phase quadrature in axially asymmetric periodic leaky-wave antennas (LWAs), so that these antennas are inherently elliptically polarized. This fact is theoretically proven and experimentally illustrated by two case-study examples, a composite right/left-handed (CRLH) LWA and a series-fed patch (SFP) LWA. Second, it shows (for the case of the SFP LWA) that the axial ratio is controlled and minimized by the degree of axial asymmetry.

## I. INTRODUCTION

THE PAST decade has witnessed a regain of interest in periodic leaky-wave antennas (LWAs) [1]–[3], due to the emergence of metamaterials and, in particular, composite right/left-handed (CRLH) transmission-line metamaterials [4], [5]. The CRLH concept has led to many novel LWA structures and systems with unique properties and functionalities [5]–[17]. Moreover, it has stimulated LWA research beyond metamaterial structures, culminating with solutions to issues which had been plaguing *conventional* LWAs for several decades, in particular the broadside issue to be discussed later.

Periodic LWAs may be implemented in waveguide, dielectric image line or planar technologies [1]–[3]. In all cases, they commonly provide high gain without requiring a complex feeding network and can be linearly or circularly polarized. For achieving circular polarization in LWAs crossed slots with an axial offset or a zigzag slot arrangement in waveguide structures [18], [19] and dielectric image lines [20], [21] have been reported. So far, CRLH LWA realizations with circular polarization mainly rely on *two* separate and linearly polarized CRLH lines fed by a hybrid coupler to achieve phase quadrature [22]–[25].

These aforementioned LWAs have common characteristics. The main beam direction can be steered from backward to forward through the critical broadside direction by increasing the frequency. When designed for moderate directivity, the power remaining at their end can be re-injected into the input via a power recycling mechanism for maximum efficiency [26], [27].

Until recently, periodic LWAs have suffered from poor radiation performance at broadside. This issue has been fully resolved for symmetric (with respect to their transversal axis) LWAs only within the past few years. The solution consists in simultaneously satisfying two distinct conditions: the closure of the stop-band at broadside and the equalization of the radiation efficiency through broadside. The stop-band closure solution was first systematically resolved by balancing the *series* and *shunt* resonances, related to *series* and *shunt radiation contributions*, respectively, in CRLH LWAs [4], [28]. The efficiency equalization was first demonstrated in CRLH LWAs, using empirical full-wave simulations, in [29]. A mathematical condition for the efficiency equalization was then provided in [30] for the specific case of a CRLH LWA. Our group has finally generalized the previous results to arbitrary periodic LWAs by introducing the concept of *Q-balancing* for achieving frequency-independent gain and efficiency when the beam is scanned through broadside [31].

To date, the polarization of the aforementioned *series radiation* and *shunt radiation* contributions has not been thoroughly examined in periodic LWAs, although the possibility to control the degree of axial asymmetry via the shunt radiation contribution has been reported for a *uniform* hybrid waveguide printed-circuit LWA in [32]. It has been speculated in [29] that the shunt radiation was cross-polarized, whereas no consideration was given to polarization in [30]. This question was addressed in [33], where it was demonstrated that shunt radiation contribution, whose amount can be controlled by the degree of axial asymmetry, is transversally polarized. At about the same time, a first simplified CRLH LWA design with *circular polarization* was presented in [34], hence demonstrating an intrinsic quadrature phase relationship between series radiation and shunt radiation for this particular LWA implementation. Consequently, a single CRLH LWA transmission line configuration was sufficient for circular polarization without the need of a hybrid coupler. Nevertheless, a theoretical treatment dealing with series and shunt radiation of arbitrary periodic LWAs has not been reported so far.

Here, we rigorously prove that LWAs composed of unit cells, which are symmetric with respect to the transversal axis and asymmetric with respect to the longitudinal axis, exhibit an intrinsic quadrature phase relationship between their series and shunt radiation contributions. Two LWA examples are demonstrated experimentally to illustrate this principle. In addition, we show that the shunt radiation contribution is controlled by the degree of longitudinal asymmetry, which may be tuned to minimize the axial ratio. We demonstrate that these characteristics can be exploited to design circularly polarized LWAs, that are topologically simple, being based on a *single* LWA line and hence do not require a hybrid coupler for feeding.

This paper is organized as follows. Section II defines the series and shunt radiation contributions and evaluates them for axially symmetric and asymmetric LWAs. Section III rigorously proves that the series and shunt radiation contributions are in quadrature phase. In Section IV, we present two case study examples, a CRLH LWA and a series fed patch (SFP) LWA, along with circuit analysis, electromagnetic simulation and measurement results to illustrate the quadrature phase relationship and the subsequent circular polarization feature. Section V next demonstrates for the SFP LWA case that the axial ratio may be controlled by the degree of asymmetry. Finally, conclusions are given in Section VI.

## II. SERIES RADIATION AND SHUNT RADIATION

This section defines and systematically derives the series and shunt radiation contributions based on voltages sampled in the LWA unit cell. This is done in preparation for Section III, where the quadrature phase relationship between the two radiation contributions will be proven.

### A. Edge Radiation from Equivalent Voltage Sources

Electromagnetic radiation follows from charge acceleration (or deceleration) [35]. In antennas, such acceleration is obtained by curving, bending, terminating, truncating or discontinuing the conductors supporting electric currents [36]. In periodic LWAs, discontinuities are the most common sources of radiation, and these discontinuities are most often provided by conductor edges or slots in conductors, which are generally modeled by magnetic current densities in antenna theory [36]. Since edge radiation and slot radiation represent similar mechanisms, by virtue of Babinet principle [35], [36], we shall focus here, without loss of generality, on edges. Radiating edges may be for instance sharp contours of metal strips in a microstrip structures [36], [37].

Fig. 1 represents the unit cell of a periodic LWA with a generic shape including several radiating edges. The overall LWA is formed by repeating this unit cell along the *y*-direction, and is assumed to be of infinite extent.[1] We furthermore assume throughout the paper that the unit cell is symmetric with respect to the transverse axis (*x*-axis) as indicated in Fig. 1.

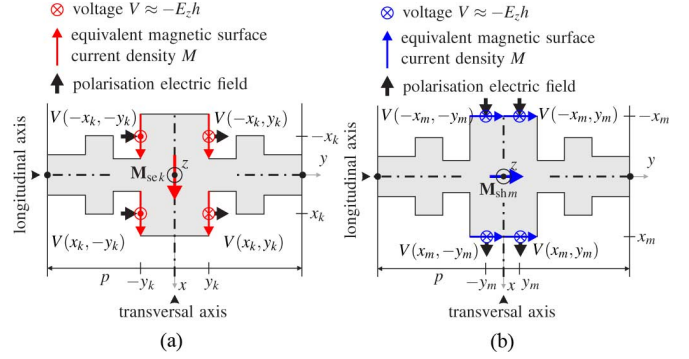

Fig. 1. Arbitrary layout of a fully symmetric unit cell with symmetric voltage sampling points $V$ at the radiating edges. For an observer placed at *broadside* in the far-field (*z*-direction), the voltage *difference* in the *y*- and *x*-directions is proportional to the series and shunt radiation field contributions, respectively. (a) The series radiation contributions (indexed $k$) from the vertical edges, polarized in the *y*-direction, are non-zero due to phase variation of the traveling wave in this direction. (b) The shunt radiation contributions (indexed $m$) from the horizontal edges, polarized in the *x*-direction, are zero due to symmetry.

The cell includes *vertical edges*, parallel to the transversal axis (*x*-axis) and supporting the magnetic currents represented in Fig. 1(a), and *horizontal edges*, parallel to the longitudinal axis (*y*-axis) and supporting the magnetic currents represented in Fig. 1(b).

We next apply the concept of the equivalent magnetic surface current density [36], to quantify the amount of radiation from the two edge types (horizontal and vertical). This current reads here

$$\mathbf{M} = \mathbf{E} \times \hat{\mathbf{n}} \approx E_z \hat{\mathbf{z}} \times \hat{\mathbf{n}} \qquad (1)$$

where $\mathbf{E}$ is the vectorial field between the conductor edge and the ground conductor, $E_z$ is the field component in the *z*-direction, and $\hat{\mathbf{n}}$ is the vector normal to the surface between the conductor edge and the ground conductor. Assuming that the edge-ground conductor distance, $h$, is much smaller than the wavelength, $h \ll \lambda/10$, $E_z$ is essentially constant, so that the voltage integral $V = \int \mathbf{E} \cdot d\mathbf{s}$ reduces to

$$V \approx -E_z h \qquad (2)$$

where the negative sign defines the voltage from the edge conductor to the ground conductor, i.e., along the negative *z*-direction.

*1) Series Radiation:* According to Fig. 1(a), the *series radiation magnetic surface current density* reads

$$\mathbf{M}_{\mathrm{se}k} = \mathbf{M}(-x_k, -y_k) + \mathbf{M}(-x_k, y_k) \\ + \mathbf{M}(x_k, -y_k) + \mathbf{M}(x_k, y_k) \qquad (3)$$

where the index $k$ refers to a sample representing any point of the vertical edges.[2] The quantity $\mathbf{M}_{\mathrm{se}k}$ measures the series radiation toward broadside and may be interpreted as a *magnetic dipole moment* located at the *center* (coordinate origin) of the unit cell with the corresponding electric far-field being polarized in the *y*-direction. If we evaluate the *series radiation magnetic*

---

[2]In Fig. 1(a), the series voltages are indicated only at the edges of the center patch for simplicity. The vertical edges of the two side patches must naturally also be taken into account, as will be seen in Fig. 3.

*surface current density* in (3) using (1) and (2) together with the proper orientation of the normal vector at the series edges, $\hat{\mathbf{n}} = \pm\hat{\mathbf{y}}$, we may write

$$\mathbf{M}_{\mathrm{se}k} \cdot \hat{\mathbf{x}} = \frac{1}{h}[-V(-x_k, -y_k) + V(-x_k, y_k) \\ - V(x_k, -y_k) + V(x_k, y_k)] \\ = \frac{1}{h}V_{\mathrm{se}k} \tag{4}$$

where the final result of $\mathbf{M}_{\mathrm{se}k}$ is projected onto the $x$-direction. The last equality in (4) defines the *series radiation equivalent voltage source*, $V_{\mathrm{se}k}$, which accounts for the series radiation contribution (amplitude and phase) as

$$V_{\mathrm{se}k} = -V(-x_k, -y_k) + V(-x_k, y_k) \\ - V(x_k, -y_k) + V(x_k, y_k) \tag{5}$$

revealing that the difference of the edge voltages in the $y$-direction contributes to series radiation, which is *y-polarized*.

*2) Shunt Radiation:* According to Fig. 1(b), the *shunt radiation magnetic surface current density* reads

$$\mathbf{M}_{\mathrm{sh}m} = \mathbf{M}(-x_m, -y_m) + \mathbf{M}(-x_m, y_m) \\ + \mathbf{M}(x_m, -y_m) + \mathbf{M}(x_m, y_m) \tag{6}$$

where the index $m$ refers to a sample representing any point of the horizontal edges. Similarly to the previous case, we use (1) and (2) together with the normal vector $\hat{\mathbf{n}}$ oriented in $\pm\hat{\mathbf{x}}$ at the horizontal edges to find the *shunt radiation magnetic surface current density* projected on the $y$-direction

$$\mathbf{M}_{\mathrm{sh}m} \cdot \hat{\mathbf{y}} = \frac{1}{h}[-V(-x_m, -y_m) - V(-x_m, y_m) \\ + V(x_m, -y_m) + V(x_m, y_m)] \\ = \frac{1}{h}V_{\mathrm{sh}m}. \tag{7}$$

The last equality in (7) defines the *shunt radiation equivalent voltage source*, $V_{\mathrm{sh}m}$, as

$$V_{\mathrm{sh}m} = -V(-x_m, -y_m) - V(-x_m, y_m) \\ + V(x_m, -y_m) + V(x_m, y_m) \tag{8}$$

showing that the edge voltage difference in the $x$-direction contributes to shunt radiation, which is *x-polarized*.

### B. Fully Symmetric Unit Cell

Consider a fully symmetric unit cell, i.e., a unit cell that is symmetric with respect to both, its longitudinal and transversal axes, as represented in Fig. 1. Assuming that the LWA is excited by a TEM field, a quasi-TEM field or by any other field whose transverse distribution is fully symmetric with respect to the longitudinal axis, i.e.,

$$V(x_k, y_k) = V(-x_k, y_k), \quad V(x_k, -y_k) = V(-x_k, -y_k). \tag{9}$$

Inserting (9) into (5) yields the equivalent series voltage

$$V_{\mathrm{se}k} = 2[-V(x_k, -y_k) + V(x_k, y_k)] \neq 0 \tag{10}$$

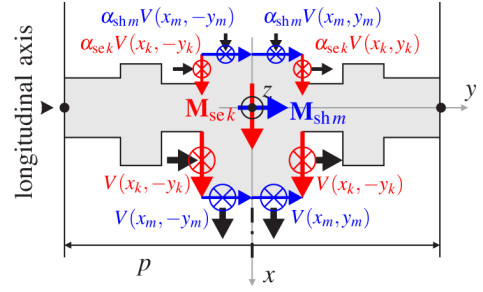

Fig. 2. Asymmetric unit cell with respect to the longitudinal axis. The series contribution, $V_{\mathrm{se}k}$, is the *difference* of the voltages with index $k$ according to (13). The shunt contribution, $V_{\mathrm{sh}m}$, is the *sum* of the voltages with index $m$. The shunt radiation is *not* canceled and therefore contributes to the far-field in the broadside direction.

which is generally nonzero since the wave propagation in the $y$-direction implies a phase difference between the two voltages $V(x_k, -y_k)$ and $V(x_k, y_k)$. Thus, the fully symmetric unit cell provides series radiation, and this radiation contribution is polarized in the longitudinal ($y$-direction.

Inserting now (9) into (8), the equivalent shunt voltage is found to be

$$V_{\mathrm{sh}m} = 0 \tag{11}$$

which reveals that no shunt radiation contribution exists, due to cancellation between the bottom edge and top edge transversally polarized ($x$) fields.

In conclusion, an LWA composed of fully symmetric unit cells radiates broadside only from its series contributions and is *linearly polarized* in its longitudinal direction.

### C. Asymmetric Unit Cell with Respect to the Longitudinal Axis

Consider now the case of a unit cell that is asymmetric with respect to the longitudinal axis while being still symmetric with respect to the transversal axis, as shown in Fig. 2. The asymmetry may be represented by *complex* asymmetry factors, $\alpha_{\mathrm{se}k}$ and $\alpha_{\mathrm{sh}m}$, for the series and shunt voltages, respectively.

Substituting

$$V(-x_k, -y_k) = \alpha_{\mathrm{se}k}V(x_k, -y_k) \tag{12a}$$

$$V(-x_k, y_k) = \alpha_{\mathrm{se}k}V(x_k, y_k) \tag{12b}$$

into (5), the *series radiation equivalent voltage source* reads

$$V_{\mathrm{se}k} = -V(x_k, -y_k)[1 + \alpha_{\mathrm{se}\,k}] + V(x_k, y_k)[1 + \alpha_{\mathrm{se}\,k}] \\ = -V_k^- + V_k^+ \tag{13}$$

where the last equality introduces the compact notation $V_k^- = V(x_k, -y_k)[1 + \alpha_{\mathrm{se}k}]$ and $V_k^+ = V(x_k, y_k)[1 + \alpha_{\mathrm{se}k}]$, the $\pm$ superscripts referring to $y$-positive/negative coordinates. The *voltage difference* of the symmetrically placed (with respect to the transversal axis) voltage samples contributes to series radiation.

Similarly, substituting

$$V(-x_m, -y_m) = \alpha_{\mathrm{sh}m}V(x_m, -y_m) \tag{14a}$$

$$V(-x_m, y_m) = \alpha_{\mathrm{sh}m}V(x_m, y_m) \tag{14b}$$

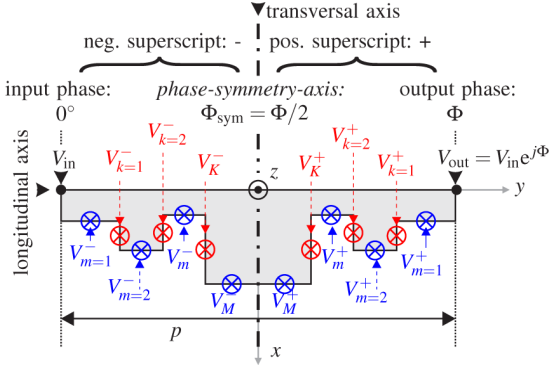

Fig. 3. Series and shunt voltage sampling at the vertical and horizontal edges with indices $k$ and $m$, respectively. In general, a phase shift $\Phi$ exists between the input/output voltages, $V_{\mathrm{in}}$ and $V_{\mathrm{out}}$, respectively. The phase along at $y = 0$, which defines the transversal axis, is therefore $\Phi_{\mathrm{sym}} = \Phi/2$ and represents the *phase-symmetry axis* of the unit cell.

into (8), the *shunt radiation equivalent voltage source* reads

$$V_{\mathrm{sh}\,m} = V(x_m, -y_m)[1 - \alpha_{\mathrm{sh}\,m}] + V(x_m, y_m)[1 - \alpha_{\mathrm{sh}\,m}] = V_m^- + V_m^+ \qquad (15)$$

where $V_m^- = V(x_m, -y_m)[1 - \alpha_{\mathrm{sh}m}]$ and $V_m^+ = V(x_m, y_m)[1 - \alpha_{\mathrm{sh}m}]$. Here, the *voltage sum* contributes to shunt radiation due to asymmetry ($\alpha_{\mathrm{sh}m} \neq 1$).

Note the simplicity of the final results of this section: series radiation is represented by a simple voltage difference [last equality in (13)], while shunt radiation is represented by a simple voltage sum [last equality in (15)]. These voltage difference and sum for the series and shunt radiation contributions may be interpreted as the odd and even components, respectively, with respect to the transversal axis of the voltages or electrical fields.

## III. PROOF OF QUADRATURE PHASE RELATIONSHIP BETWEEN THE SERIES AND SHUNT FIELDS

Based on the simple odd/even difference/sum formalism established in Section II, this section proves that the series and shunt radiation contributions, $V_{\mathrm{se}k}$ and $V_{\mathrm{sh}m}$, are in quadrature phase for arbitrary $k$ and $m$.

Fig. 3 shows an LWA unit cell that is asymmetric with respect to the longitudinal axis with the index $k$ accounting for the vertical edges and the index $m$ accounting for the horizontal edges. For visualization simplicity, Fig. 3 represents the case $\alpha_{\mathrm{se}k} = 0$ and $\alpha_{\mathrm{se}m} = 0$.[3]

Setting a phase reference of $0°$ at the input of the unit cell (at $y = -p/2$) and denoting the phase shift due to wave propagation across the unit cell $\Phi$, the phase at the output of the unit cell (at $y = +p/2$) is $\Phi$, and the phase on the transversal axis, at the center of the unit cell (at $y = 0$), is

$$\Phi_{\mathrm{sym}} = \frac{\Phi}{2}. \qquad (16)$$

The transversal axis is thus the *phase-symmetry axis* of the unit cell. Here, it is assumed that the unit cell is: 1) topologically

---

[3]The quadrature phase relationship is independent of the $\alpha_{\mathrm{se}k}$ and $\alpha_{\mathrm{sh}m}$ coefficients, whose importance is discussed later in this section.

symmetric with respect to the transverse axis; 2) lossless;[4] 3) that the LWA operates in the passband.[5]

The complex vertical (series) and horizontal (shunt) edge voltages at the points $k$ and $m$, respectively, may be written as

$$V_k^\pm = |V_k^\pm| e^{j\Phi_k^\pm}, \qquad V_m^\pm = |V_m^\pm| e^{j\Phi_m^\pm}. \qquad (17)$$

By symmetry about the phase-symmetry axis, one may write

$$V_k^+ = |V_k^+| e^{j(\Phi_{\mathrm{sym}} + \Delta\Phi_k)}, \qquad V_k^- = |V_k^-| e^{j(\Phi_{\mathrm{sym}} - \Delta\Phi_k)} \qquad (18a)$$

where

$$\Delta\Phi_k = \frac{(\Phi_k^+ - \Phi_k^-)}{2} \qquad (18b)$$

is the phase difference between $\Phi_{\mathrm{sym}}$ and $\angle V_k^\pm$.

Similarly, the shunt voltages follow with

$$V_m^+ = |V_m^+| e^{j(\Phi_{\mathrm{sym}} + \Delta\Phi_m)}, \qquad V_m^- = |V_m^-| e^{j(\Phi_{\mathrm{sym}} - \Delta\Phi_m)} \qquad (19a)$$

where

$$\Delta\Phi_m = \frac{(\Phi_m^+ - \Phi_m^-)}{2} \qquad (19b)$$

is the phase difference between $\Phi_{\mathrm{sym}}$ and $\angle V_m^\pm$.

Inserting (18a) into (13) yields

$$V_{\mathrm{se}k} = |V_k^+| e^{j(\Phi_{\mathrm{sym}} + \Delta\Phi_k)} - |V_k^-| e^{j(\Phi_{\mathrm{sym}} - \Delta\Phi_k)} \qquad (20)$$

for the series radiation equivalent voltage source, while inserting (19a) into (15) yields

$$V_{\mathrm{sh}m} = |V_m^+| e^{j(\Phi_{\mathrm{sym}} + \Delta\Phi_m)} + |V_m^-| e^{j(\Phi_{\mathrm{sym}} - \Delta\Phi_m)} \qquad (21)$$

for the shunt radiation equivalent voltage source.

We now recall that the structure is lossless, having neither radiation loss nor dissipation loss.[4] Although this assumption goes against the physical nature of an LWA, it serves as a reasonable approximation for the purpose of the forthcoming argument because the amount of leakage and dissipation across a periodic LWA unit cell is very small [1]–[3]. Under this assumption, we have, a fortiori for points within the unit cell

$$|V_k^+| = |V_k^-| = |V_k|, \qquad |V_m^+| = |V_m^-| = |V_m|. \qquad (22)$$

Under these conditions, (20) and (21) reduce, via Euler's formula, to the expressions

$$V_{\mathrm{se}k} = 2j|V_k| \sin(\Delta\Phi_k) e^{j\Phi_{\mathrm{sym}}} \qquad (23)$$

$$V_{\mathrm{sh}m} = 2|V_m| \cos(\Delta\Phi_m) e^{j\Phi_{\mathrm{sym}}} \qquad (24)$$

whose ratio

$$\frac{V_{\mathrm{se}k}}{V_{\mathrm{sh}m}} = j \frac{|V_k| \sin(\Delta\Phi_k)}{|V_m| \cos(\Delta\Phi_m)} \qquad (25)$$

---

[4]Typically, $\alpha/k_0 = \alpha\lambda_0/(2\pi) \sim 10^{-2}$, where $\alpha$ is the leakage factor, $k_0$ is the free-space wavenumber and $\lambda_0$ is the free-space wavelength, with $p \lesssim \lambda_0$, so that $\alpha p \lesssim \alpha\lambda_0 \sim 2\pi \cdot 10^{-2}$. Then the power lost across the unit cell due to leakage is $P_{\mathrm{u.c.}}^{\mathrm{leakage}} = |20\log_{10} e^{-\alpha p}| < 0.5$ dB ($e^{-\alpha p} > 0.94$), while the corresponding power due to dissipation is usually much smaller than this quantity at microwaves.

[5]For operation in the passband the phase angle $\Phi$ across the unit cell is a *real* quantity, as opposed to an operation in the stop-band of the LWA, where $\Phi$ would be purely *imaginary*.

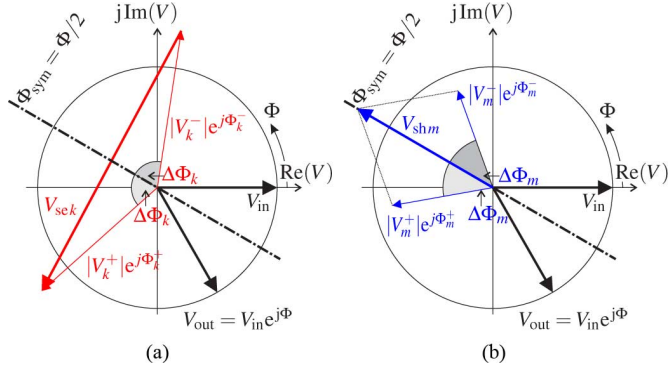

Fig. 4. Graphical proof of the quadrature phase relationship between $V_{se\,k}$ and $V_{sh\,m}$ in the complex voltage plane. Both graphs include the voltages at the input and at the output of the unit cell, $V_{in}$ and $V_{out}$, respectively, and the phase-symmetry axis (dashed line). The vectors $V_{se\,k}$ and $V_{sh\,m}$, obtained by the difference and sum of the corresponding series or shunt sample vectors according to (20) and (21), respectively, and fulfilling the condition (22), are perpendicular to each other, and thus in phase quadrature. (a) $V_{se\,k}$. (b) $V_{sh\,m}$.

proves the quadrature phase relationship between the series and shunt radiation contributions. This holds for unit cells that are symmetric with respect to their transverse axis and asymmetric with respect to their longitudinal axis.

$V_{se\,k}$ and $V_{sh\,m}$ are generally non-zero. In the particular case of a CRLH LWA, it might a priori seem that $V_{se\,k}$ in (20) vanishes because at broadside the input/output phase shift is zero, $\Phi = 0$, so that $\Phi_{sym} = 0$ according to (16). On the other hand $\Delta\Phi_k$ and $\Delta\Phi_m$ are non-zero due to the voltage variations occurring *within* the unit cell [38]. So, (25) is clearly determined and provides the axial ratio of the polarization.

Fig. 4 provides a graphical representation of this phase quadrature proof. Fig. 4(a) plots the vector $V_{se\,k}$ as the *difference* of the vectors $V_k^+$ and $V_k^-$, according to (20) with (18), while Fig. 4(b) plots the vector $V_{sh\,m}$ as the *sum* of the vectors $V_m^+$ and $V_m^-$, according to (21) with (18). In both of these graphs, the symmetries about the phase-symmetry axis between the input and output voltages and between the negative and positive point voltages is clearly apparent, and the proof follows from the perpendicularity between the complex numbers $V_{se\,k}$ and $V_{sh\,m}$

$$V_{se\,k} = \xi V_{sh\,m}\,e^{j\pi/2} = j\xi V_{sh\,m} \tag{26a}$$

or

$$\frac{V_{se\,k}}{V_{sh\,m}} = j\xi \tag{26b}$$

which provides the same quadrature phase relationship as (25), with the real proportionality factor

$$\xi_{km} = \frac{|V_k|\sin(\Delta\Phi_k)}{|V_m|\cos(\Delta\Phi_m)}. \tag{27}$$

Since it has been demonstrated for *arbitrary* voltage sample points ($k$ and $n$), phase quadrature naturally extends to the *integral sum* of all the edge voltages associated with radiation. Therefore, the phase quadrature proof provided above applies to the *entire* LWA structure, and reveals that an LWA whose unit cell is asymmetric with respect to its longitudinal axis necessarily exhibits elliptical polarization radiation, or *circular polarization radiation* with broadside axial ratio obtained from (27) along with (13) and (15) as

$$\xi = \frac{\Sigma_k |V_k(\alpha_{se\,k})|\sin(\Delta\Phi_k)}{\Sigma_m |V_m(\alpha_{sh\,m})|\cos(\Delta\Phi_m)}. \tag{28}$$

It is therefore expected that the axial ratio of the LWA can be controlled by tuning the amount of asymmetry with respect to the longitudinal axis, as done in [32] for the case of an *uniform* LWA. This will be demonstrated in Section V.

Note that only the broadside case has been considered in the analyses of this section and the previous one. However, as will be shown in Section IV, the conclusions of these analyses also essentially apply to off-broadside radiation directions within a scanning range in the order of $[-20°, +20°]$ around broadside. Let us investigate why this is the case by inspecting the fundamental relation (25). In terms of phase, (25) is inherently broadband, since the phase is always $\pi/2$ irrespectively to frequency. So, the conclusions for broadside are valid also at off-broadside angle as far as quadrature phase is concerned. In contrast, in terms of amplitude, (25) clearly depends on frequency via $k$, $m$. However, when the radiating array structure is electrically relatively large, as is the case in LWAs, the radiation characteristics are essentially determined by the array factor while the element factor plays a minor role [39]. So, since the frequency sensitivity of the unit cell magnitude in (25) is an element factor quantity, it can only have a minor effect on the radiation property of the LWA. Hence, despite its frequency dependence, (25) is a broadband relationship and can thus extend to off-broadside radiation angles.

## IV. EXAMPLES: CRLH LWA AND SFP LWA

This section provides two examples of longitudinally-asymmetric LWAs to illustrate the quadrature phase relationship derived in the previous section, a CRLH LWA and an SFP LWA. First, the quadrature phase relationship is illustrated by comparing voltage samples of an ideal circuit models with corresponding samples obtained from full-wave simulation. Then, simulated and measured far-field pattern and axial ratio results are presented.

### A. CRLH and SFP LWA Structures

Fig. 5(a) and (b) show the unit cell geometries and layer stack-up for the CRLH and SFP LWAs, respectively. The corresponding geometry data is provided in Table I.

Both LWAs are optimized to exhibit their transition frequency (frequency where they radiate broadside) at 24 GHz. Around this broadside frequency, they can be modeled by means of equivalent series and shunt resonators of resonance frequency $f_{se} = f_{sh} = f_0$, [31], [33]. In the CRLH LWA, $l_g$ is used to tune $f_{se}$ and $l_s$ is used to tune $f_{sh}$. In the SFP, we have to tune both $l_1$ and $l_p$ simultaneously, since each of them affects $f_{sh}$ and $f_{se}$, [31]. Whereas the resonators are formed by quasi-lumped elements ($p \ll \lambda$) in the CRLH LWA, in the SFP LWA the resonators are formed by distributed (transmission-line) resonators with an overall length of $p \approx \lambda$ across the unit cell. Hence, the CRLH LWA is operating in its fundamental

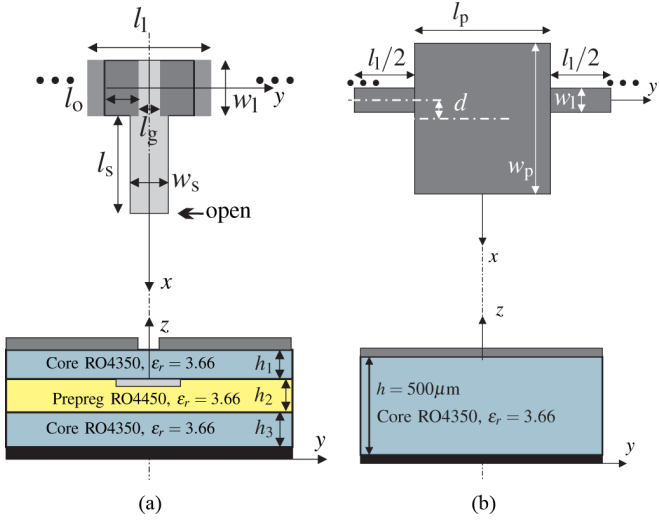

Fig. 5. Examples of periodic LWAs with a longitudinally-asymmetric unit cell. Shown are the layout (top) and layer stack-up (bottom). (a) Unit cell of a CRLH LWA with metal insulator metal (MIM) capacitor ($h_1 = 170\ \mu m$, $h_2 = 100\ \mu m$, $h_3 = 250\ \mu m$). (b) Unit cell of a series fed patch (SFP) LWA.

TABLE I
LAYOUT DIMENSIONS ($\mu$m) FOR THE CRLH UNIT CELL IN FIG. 5(a)
AND THE SFP UNIT CELL IN FIG. 5(b)

| CRLH | | SFP | |
|---|---|---|---|
| $l_l$ | 2250 | $l_l$ | 3200 |
| $w_l$ | 700 | $l_p$ | 3425 |
| $l_g$ | 350 | $w_l$ | 300 |
| $l_o$ | 450 | $w_p$ | 1800 |
| $w_s$ | 470 | $d$ | 750 |
| $l_s$ | 1875 | | |

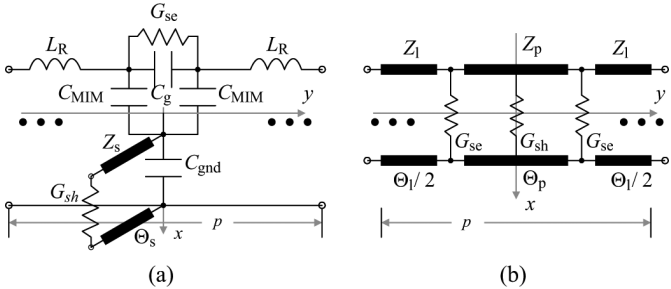

Fig. 6. Circuit models of the unit cells in Fig. 5. (a) CRLH LWA. (b) SFP LWA.

space harmonic ($n = 0$), while the SFP is using its first space harmonic ($n = -1$) [31].

### B. Circuit Modeling

Fig. 6(a) and (b) show the equivalent circuit models for the CRLH LWA and the SFP LWA, respectively.

The CRLH LWA circuit model comprises the following circuit elements: the series inductance of the small transmission-line section $L_R$, the gap capacitance $C_g$, the capacitance to the middle metalization layer $C_{MIM}$, the ground capacitance $C_g$, and the transmission line of impedance $Z_s$ and length $l_s$ for the open ended stub line. For completeness, we also show the se-ries and shunt radiation conductances in the gap and at the end of the stub line, respectively.

TABLE II
ELEMENT PARAMETERS FOR THE EQUIVALENT CIRCUIT OF THE CRLH UNIT
CELL IN FIG. 6(a) AND THE SFP UNIT CELL IN FIG. 6(b)

| CRLH | | SFP | |
|---|---|---|---|
| $L_R$ | 0.44 nH | $Z_l$ | 90 $\Omega$ |
| $C_g$ | 0.02 pF | $Z_p$ | 35 $\Omega$ |
| $C_{MIM}$ | 0.06 pF | $\Theta_l$ | 180° |
| $C_{gnd}$ | 0.43 pF | $\Theta_p$ | 180° |
| $Z_s$ | 40.5 $\Omega$ | | |
| $\Theta_s$ | 110.82° | | |

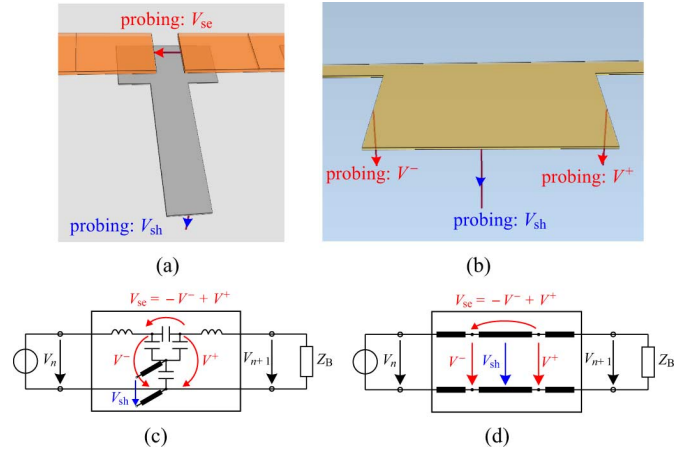

Fig. 7. Voltage probing across the unit cell in a periodic structure (infinite repetition of the unit cell). (a) and (b) show the electromagnetic models with voltage probes for the CRLH and the SFP unit cell, respectively. A quasi-periodic environment is emulated by including a large number of unit cells so minimize reflections ($< -40$ dB) from the end. (c) and (d) show the *lossless* circuit models with voltage probes for the CRLH and the SFP unit cell, respectively. A perfectly periodic environment is emulated by terminating the unit cell with the exact, *a priori* calculated, Bloch impedance $Z_B$.

The SFP LWA is modeled by three transmission-line sections, with different impedance, of: 1) quarter wavelength; 2) half wavelength; and 3) quarter wavelength at the broadside frequency. The quarter-wavelength lines have the impedance $Z_l$ and the half-wavelength patch has the impedance $Z_p$. Series radiation is modeled by the conductance $G_{se}$ located at the vertical patch edge and the shunt conductance $G_{sh}$ located at the center of the horizontal patch edge.

Table II provides the extracted circuit parameters, obtained through a curve fitting procedure from full-wave simulated scattering parameters using the circuit simulator of ADS. The conductances modeling radiation have been set to infinity to represent a lossless circuit model under the same conditions as the theoretical proof in Section III.

### C. Quadrature Phase Relationship

We now provide numerical results from circuit and full-wave simulations to illustrate the established quadrature phase relationship proven in Section III.

Fig. 7(a) and (b) show the full-wave models in EMPIRE XCcel, a commercial FDTD-based simulation tool, with the voltage probing positions. The voltages are obtained from electric field integration in $z$-direction. A sufficient number of $n = 100$ unit cells have been cascaded and the voltages have been sampled at the center cell ($n = 50$), to avoid reflection and truncation effects associated with the finiteness of the structure.

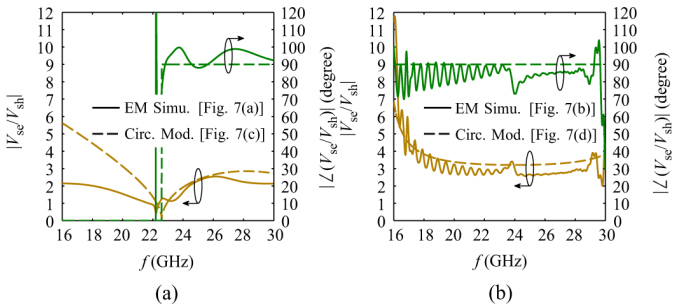

Fig. 8. Voltage ratio $V_{se}/V_{sh}$ amplitude and phase obtained from circuit (dashed lines) and full-wave simulations (solid lines). A broadband phase difference of $90°$ between $V_{se}$ and $V_{sh}$ is observed in an infinite periodic configuration for both LWAs. (a) CRLH unit cell. The amplitude and phase evaluation below approximately 22 GHz (lower stop-band) is not meaningful. (b) SFP unit cell having a broadband quadrature phase relationship of 14 GHz.

Fig. 7(c) and (d) show the corresponding circuit models. A truly infinite periodic environment is emulated by terminating the unit cell with the exact, *a priori* calculated Bloch impedance, $Z_B$, coinciding with the input impedance for an infinite structure. This has been done through a two-step procedure: first, the Bloch impedance was calculated as $Z_B = \sqrt{B/C}$ ($ABCD$ is the transfer matrix with $A = D$ due to sym. with respect to the transversal axis); next, a second simulation was launched, sampling the voltages, with the unit cell driven by a voltage source and terminated with $Z_B$. The circuit modeling is essentially lossless and therefore does not account for radiation. This is in order to follow the assumption used in the proof of Section III.

The voltage ratio, $V_{sek}/V_{shm}$, as given by (25), is plotted in Fig. 8, which compare circuit and full-wave simulation results.

Fig. 8(a) plots the voltage ratio amplitude and phase for the CRLH LWA. A perfect broadband phase quadrature is clearly confirmed by the circuit simulation result. The CRLH structure, whose transition frequency is at 24 GHz, has its LH low-pass *stop-band* [4] at around 22 GHz. Below this frequency, the amplitude and phase evaluation is naturally meaningless. The full-wave results are consistent with the circuit ones, and support the phase quadrature within a 10% variation. The ratio of the amplitude varies from 0 at the lower stop band to about 3 at 30 GHz, which will affect the axial ratio over frequency.

Fig. 8(b) plots the voltage ratio amplitude and phase for the SFP LWA. The circuit simulation reveals an ideal quadrature phase relationship from 16 GHz to 30 GHz, covering a huge bandwidth of 14 GHz. Across this bandwidth, the phase variation is only 20% in the full-wave model, which will correspond to an axial ratio limit for circular polarization. The amplitude, which is the second quantity of interest for obtaining a small axial ratio, is almost constant in the range between 18 and 28 GHz, hence promising a rather frequency-independent axial ratio.

The phase deviation of the electromagnetic simulation results from the theoretical results in Fig. 8 is due to radiation loss, air coupling between unit cells and to the existence of a small band gap in the SFP case in Fig. 8(b), effects which are not taken into account in theory (lossless circuit simulation). Reflections from the end of the finite LWA were minimized by using a large number of cascaded cells, so one can be sure that this phase deviation is *not* attributed to reflections from the LWA end.

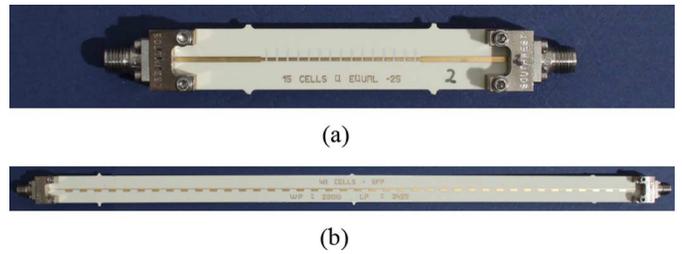

Fig. 9. Fabricated LWA prototypes with end-feed connectors. (a) CRLH LWA with 15 unit cells (electrical length $\approx 2.8\lambda$). (b) SFP LWA with 40 unit cells (electrical length approximately $\approx 22\lambda$).

In summary, Fig. 8 verify the broadband nature of the quadrature phase relationship required as a first condition for circular polarization. The second condition—equal field amplitudes—is not exactly represented by sampled voltages, since accurate determination of the radiated far-field requires integration over all radiating edges.

Therefore, the amplitudes in Fig. 8 are *relative* quantities in terms of the radiated fields, which only indicate the expected frequency variation of the axial ratio.

### D. Experimental Validation

Fig. 9(a) and (b) show the fabricated prototypes for the CRLH and SFP LWAs, respectively. The CRLH LWA is composed of 15 unit cells and is electrically relatively small, featuring a length of $L \approx 2.8\lambda$ without the feeding-line sections. The SFP LWA is electrically much larger, with $L \approx 22\lambda$ compared to the CRLH LWA. The reason for the different electrical lengths is simply that these antennas were fabricated in different contexts. They are naturally not comparable in size, but the size difference does not represent any limitation to the illustration to be presented next, since the polarization does not depend on the length of the LWA.

Whereas the CRLH Bloch impedance is in the order of $50\,\Omega$, hence requiring no additional matching, the SFP requires a quarter-wavelength transformer at its input to match the Bloch impedance of $\sim 170\,\Omega$ to the $50\,\Omega$ external ports. The LWAs are excited from one side while the other side is terminated with a matched load.

An enlarged view of the prototypes is provided in Fig. 10. The dashed lines in Fig. 10(a) indicate the middle metalization layer beneath the top metalization.

The LWAs have been characterized by a cylindrical near-field measurement in the anechoic chamber of the company IMST GmbH, shown in Fig. 11. The electric field is sampled on a cylindrical surface (excluding the circular top and bottom faces, where radiation is negligible) using an open-ended wave guide probe. The vertical and horizontal fields are measured, so as to fully characterize the LWA in their near-field zone. Finally, the near-field data are transformed to the far-field using the commercial antenna measurement software package MiDAS (from ORBIT/FR, Inc) in order to provide the antenna gain and polarization.

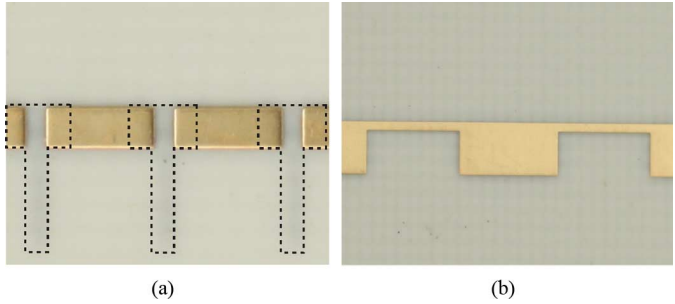

Fig. 10. Enlarged views of the fabricated LWAs showing the unit cell details. (a) CRLH LWA. (b) SFP LWA.

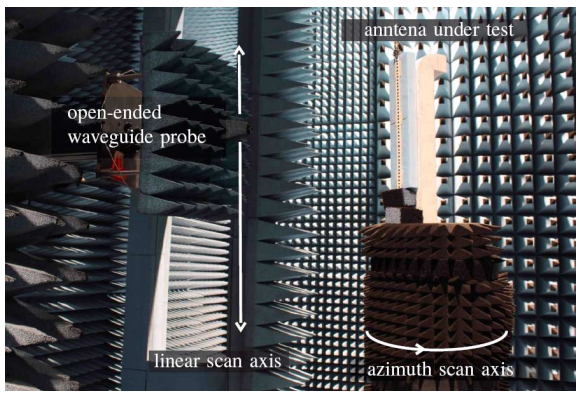

Fig. 11. Cylindrical near-field measurement setup in anechoic chamber. The measured near-field data are transformed to the far-field and the RHCP/LHCP gain components are evaluated using the commercial measurement software MiDAS.

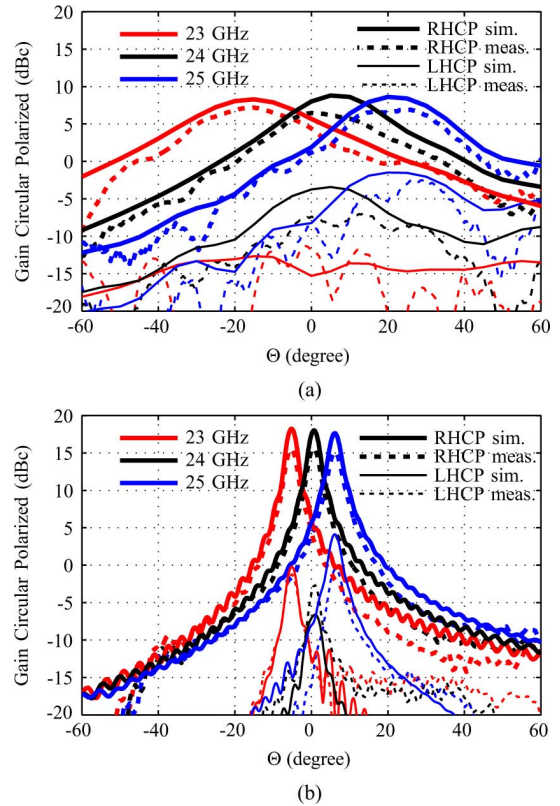

Fig. 12. Measured and simulated radiation patterns in the scanning plane. Circular polarization is confirmed by comparing RHCP (co-pol.) gain and LHCP gain (x-pol.) for three frequencies 23, 24, and 25 GHz. (a) CRLH LWA. (b) SFP LWA.

Fig. 12 show the simulated and measured co- and cross-polarization gains at the frequencies 23, 24, and 25 GHz. Both LWAs are right-hand circular polarized (RHCP) with the left-hand circular polarization (LHCP) being the cross polarization. In general we have an excellent agreement between the simulated and measured co-polarization, where as the agreement of the cross polarization is still acceptable.[6]

Fig. 12(a) shows that the CRLH LWA exhibits a relatively large half-power beamwidth, as an expected consequence of its small electrical length. On the other hand the LWA scanning sensitivity with frequency is relatively high, due to the small period $p$ of the unit cell [31]. This LWA has a moderate circular polarization response, with a cross polarization rejection varying from about 23 dB at 23 GHz up to 10 dB at 25 GHz, as shown in Fig. 13(a). The agreement between simulation and measurement is reasonably good in the main beam direction. Comparison of the axial ratio in Fig. 13(a) with the voltage-based amplitudes in Fig. 8(a) shows the same trend of increase with frequency for the CRLH case.

Unfortunately, only the backward radiation region, at 23 GHz, exhibits a good axial ratio, of less than 3 dB. However,

no optimization has been carried out, and this result experimentally demonstrates the circular (or elliptical) polarization response of the longitudinally-asymmetric CRLH LWA.

Fig. 12(b) shows that the SFP LWA exhibits a smaller half-power beamwidth, as a result of its electrically much larger size. The scanning sensitivity with frequency of this LWA is much smaller than that of the SFP, due to its larger period $p$. This LWA has an excellent circular polarization response at broadside, with a cross polarization rejection of more than 20 dB. For the backward and forward radiation zones, the cross polarization rejection is less, ranging from 18 to 13 dB. The corresponding axial ratio is plotted in Fig. 13(b), consistently with the cross polarization rejection in Fig. 12(b). Comparison of the axial ratio in Fig. 13(b) with the voltage based amplitudes in Fig. 8(b) shows the same frequency insensitivity for the SFP case. So, again without any optimization, the circular polarization characteristics of the SFP LWA have been experimentally demonstrated.

## V. AXIAL RATIO OPTIMIZATION

This section presents a parametric study showing that, as predicted by Section III, the axial ratio of longitudinally-asymmetric LWAs can be minimized by adjusting the degree of asymmetry of the structure, as done in [32]. This study is presented only for the case of the SFP LWA, whose asymmetry does not affect the resonance frequencies. In the CRLH LWA, the stub length controls the shunt resonance frequency and can therefore not be exploited as an independent asymmetry parameter.

---

[6]The cross polarization is more sensitive to fabrication, alignment and measurement errors and the resolution is limited by the dynamic range of the measurement system.

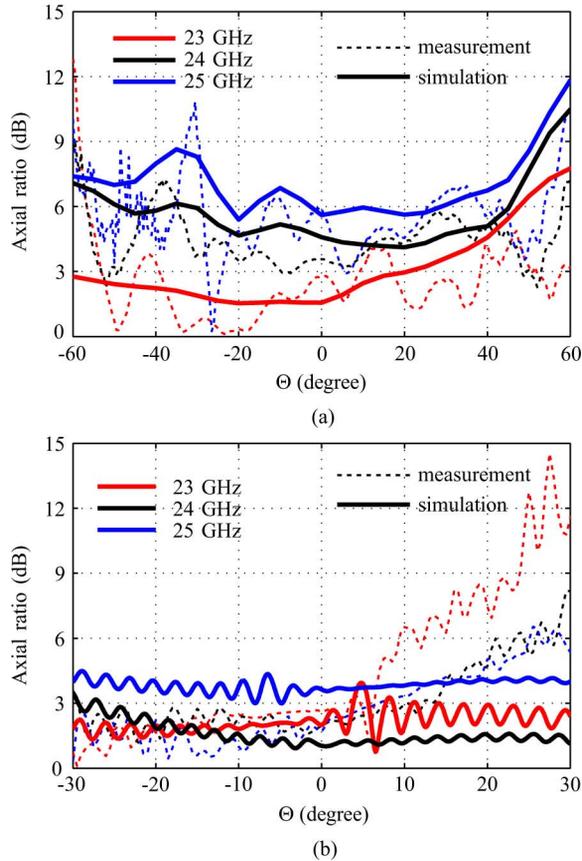

Fig. 13. Axial ratio (dB) in the scanning plane. (a) CRLH LWA. (b) SFP LWA.

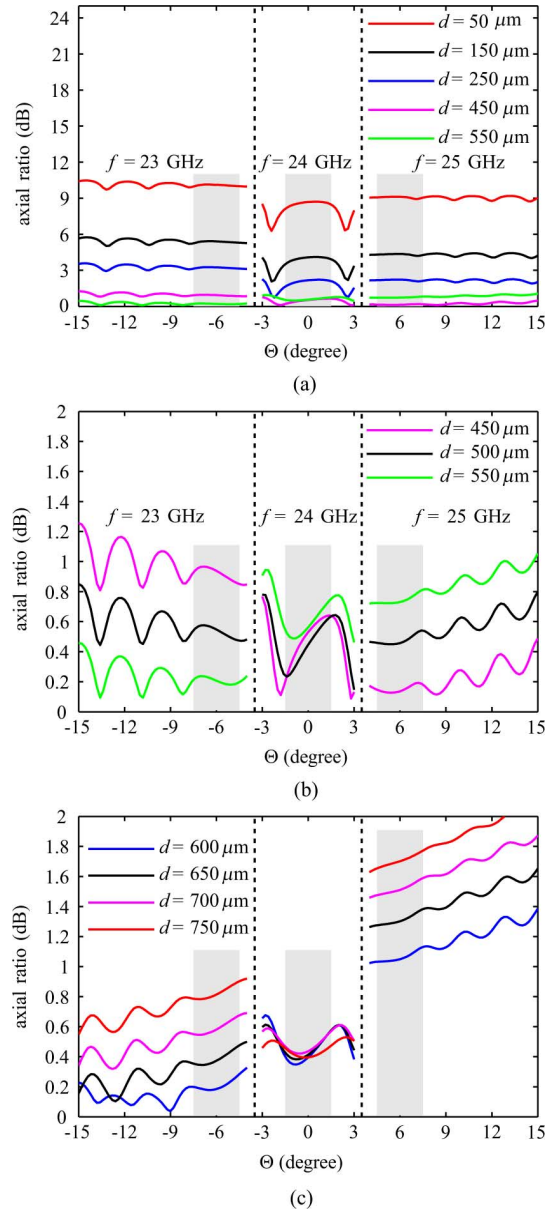

Fig. 14. Simulated axial ratio (dB) in the scanning plane using EMPIRE XCcel (FDTD). The degree of asymmetry with respect to the longitudinal axis is parameterized by $d$. The shaded areas indicate the half-power beamwidth. (a) Coarse variation from $d = 50\ \mu m$ to $d = 550\ \mu m$. (b) Fine variation with ±50 $\mu m$ around the optimum axial ratio of $d = 500\ \mu m$. (c) Fine variation with ±50 $\mu m$ from $d = 550\ \mu m$ to $d = 750\ \mu m$, which represents the maximum asymmetry, see prototype in Fig. 10(b).

The degree of asymmetry is proportional to the parameter $d$, shown in Fig. 5(b), which represents an offset between the longitudinal axis of the feeding line and the center of the patch. For each $d$ value, the line length ($l_1$) and the patch length ($l_p$) are optimized to achieve broadside radiation at $f_0 = f_{se} = f_{sh} = 24$ GHz from equalized series and shunt resonances.

Fig. 14(a) plots the SFP LWA axial ratio in the scanning plane for different values of $d$ versus the elevation angle for three frequencies, 23, 24, and 25 GHz, corresponding to the beam pointing angles $-6°$, $0°$, and $+6°$, respectively. The elevation angular range was divided in three sectors: $-15°$ to $-4°$ for backward radiation (23 GHz), $-3°$ to $+3°$ for broadside radiation (24 GHz) and $+4°$ to $+15°$ for forward radiation (25 GHz) to evaluate the axial ratios of the main beams. The half-power beamwidth in each sector is indicated by the shaded rectangular zones. It clearly appears that the axial ratio decreases with increasing $d$ up to $d = 450\ \mu m$. Fig. 14(b) provides the axial ratio over a smaller scale for better visualization of the results around $d = 500 \mu m$. The best tradeoff, with an axial ratio of less than 0.8 dB in the backward, broadside and forward regions, is found at $d = 500\ \mu m$. If $d$ is further increased as done in Fig. 14(c), we observe in the backward and forward regimes that the trend is reversed, the axial ratio is increasing with further increasing asymmetry.

Fig. 14(b) and (c) show, for asymmetries providing *near optimum* axial ratios, that the axial ratio at broadside is: 1)

relatively low and close to 0 dB and 2) relatively insensitive to asymmetry variation, whereas in the backward and forward regimes it strongly depends on asymmetry variation. The low axial ratio at broadside and the sensitivity difference of the axial ratio between the broadside and off-broadside regimes will next be explained by investigating the series and shunt powers, $P_{se}$ and $P_{sh}$, in the two regimes [31].

In the forthcoming developments, we neglect dissipation loss, which is a reasonable assumption in typical cases where radiation leakage is significantly larger than dissipation. This is equivalent to assuming that the total series and shunt powers are equal to the radiated series and shunt powers. In addition, we now restrict to axial asymmetries which provide *near optimum* axial ratios. Radiation at broadside results from the sum of two equivalent and orthogonal magnetic dipoles representing respectively the series and shunt radiation contributions, as indicated in Fig. 2. Although rotationally offset by 90° in their common plane ($xy$-plane), these dipoles, given their identical toroidal radiation patterns, exhibit relatively similar radiation patterns around broadside. Assuming that these patterns are identical, which is a reasonable approximation in a small solid angle around broadside, including small backward and forward angles, the axial ratio can be simply expressed in terms of the series and shunt powers.[7] Under these three conditions (no dissipation, near-broadside radiation and *near-optimum* axial ratios), the axial ratio is simply

$$AR \approx \sqrt{\frac{P_{se}}{P_{sh}}}. \tag{29}$$

*1) Broadside:* According to [31], exactly at broadside and under the frequency-balanced condition ($\omega_{se} = \omega_{sh} = \omega_0$), the LWA structure is modeled by purely *resistive* circuit elements and therefore, the total series and shunt powers (radiation + dissipation) are always equal,[8] i.e.,

$$P_{se} = P_{sh} = P_{at-BS} \tag{30}$$

where again dissipation is assumed to be zero. Inserting (30) into (29) leads to

$$AR_{at-BS} \approx 1. \tag{31}$$

This result shows that, under the aforementioned conditions, the axial ratio exactly at broadside is always equal to one (0 dB), corresponding to perfect circular polarization. Second, being a constant (equal to one), the axial ratio is obviously invariant under asymmetry variation for near-optimum values.

*2) Off-Broadside:* The quality factor of an antenna is generally defined as [40], [41], [42]

$$Q = \omega \frac{2\max(W_E, W_M)}{P} = \omega_0 \frac{W}{P}\bigg|_{\omega_0} \tag{32}$$

where $W = W_E + W_M$ is the maximum stored energy, with the electric energy, $W_E$, and the magnetic energy, $W_M$, being equal at the resonance frequency, $\omega_0$, and $P$ is the radiated power. In the off-broadside regime and under the frequency-balanced condition, the series and shunt powers are then related to their respective quality factors by

$$P_{se} = \omega_0 \frac{W_{se}}{Q_{se}} \quad \text{and} \quad P_{sh} = \omega_0 \frac{W_{sh}}{Q_{sh}}, \tag{33}$$

where $W_{se}$, $W_{sh}$ and $Q_{se}$, $Q_{sh}$ are the series and shunt energies and quality factors, respectively. Note that the energies model the series and shunt reactive near-fields of the LWA and are not related to radiation.

According to [31], off-broadside the LWA structure is modeled by purely *reactive* circuit elements and therefore, the series and shunt energies are always equal, i.e.,

$$W_{se} = W_{sh} = W_{off-BS}. \tag{34}$$

Substituting (34) into (33), and inserting the result into (29) leads to

$$AR_{off-BS} \approx \sqrt{\frac{Q_{sh}}{Q_{se}}}. \tag{35}$$

This result shows that the axial ratio is generally not equal to one and that it varies according to the ratio of the series and shunt quality factors for near-optimum values. Moreover, (35) reveals that, although the off-broadside axial ratio is generally different from one, it is exactly one in the particular case of $Q$-balancing ($Q_{se} = Q_{sh}$) [31], which is thus also the condition for minimum axial ratio off-broadside. From a practical viewpoint, this last point is very convenient, since it indicates that optimum circular polarization is inherently associated with optimum broadside radiation.

Since the axial ratio in (35) depends on the two quality factors, we need to investigate the dependence of the quality factors on asymmetry to better understand the corresponding axial ratio behavior. The series and shunt quality factors of the SFP LWA are plotted in Fig. 15 versus the asymmetry parameter $d$ as defined in Fig. 5(b). These quality factors were obtained using the driven-mode extraction technique (from full-wave simulation) described in [31].

Fig. 15 shows that $Q_{sh}$ strongly depends on asymmetry, whereas $Q_{se}$ is essentially insensitive to asymmetry. $Q_{sh}$ is maximum for the fully symmetric configuration ($d = 0$), where shunt radiation is canceled at broadside, and strongly decreases when asymmetry is introduced and increased, according to (33), due to the progressive reduction of radiation cancellation associated with increased shunt radiation power. The asymmetry invariant behavior of $Q_{se}$ is due to the absence of structural variation along the longitudinal axis.[9] The $Q$-balancing condition is approximately met at $d \approx 550$ $\mu$m.

The axial ratio behavior in Fig. 14 may now be understood by evaluating (31) at broadside, and (35) together with the $Q$-factors in Fig. 15 off-broadside.

In Fig. 14(a) the axial ratio at broadside does *not* behave according to (31), where a constant and low axial ratio, independent of asymmetry, is expected. The reason is that the shunt radiation pattern is strongly affected by the asymmetry in near-symmetric structures. Indeed, the shunt radiation pattern has a null at broadside for the fully symmetric configuration and its pattern is thus totally different from the series radiation pattern, which exhibits a maximum at broadside. Here, the aforementioned assumption of identical patterns is thus violated, and hence the behavior of the axial ratio is not captured by (31). On the other hand, the axial ratio in the off-broadside regime qualitatively behaves according to (35), i.e., it decreases as $Q_{sh}$ decreases.

---

[7]We have seen that in the case of a fully symmetric structure, there is a null in the shunt pattern at broadside, leading to linear longitudinal polarization. So, this argument naturally relates only to the case of asymmetric structures, in particular *near optimum* axial ratio structures.

[8]This is true for both axially symmetric and axially asymmetric LWAs, as long as they are symmetric with respect to the transversal axis.

[9]The parameter $d$ does not change the overall length of the series radiating edges, which remains constant: $w_p - w_l = 1500$ $\mu$m.

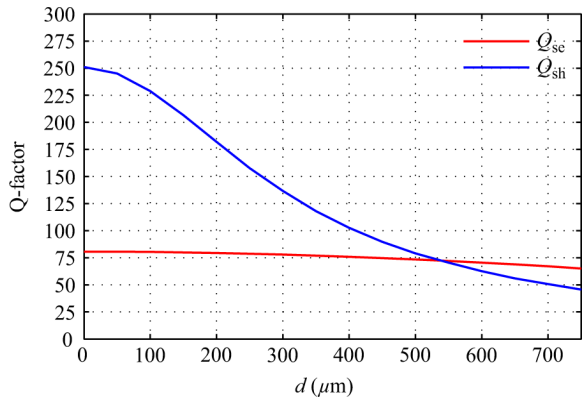

Fig. 15. Simulated series and shunt quality factors of the unit cell versus the patch offset $d$ (degree of longitudinal asymmetry). $Q$-balancing is achieved at $d \approx 550\ \mu m$.

Fig. 14(b) and (c) show the axial ratio in the near-optimum asymmetry range. Here, the radiation characteristics of the series and shunt contributions are not strongly affected by the asymmetry variation, due to the similarity of the corresponding radiation pattern. At broadside, the axial ratio is relatively low and insensitive to asymmetry variations, in agreement with (31). Off-broadside, the exact behavior predicted by (35) is observed, with $Q_{se}$ being constant and $Q_{sh}$ decreasing with increasing $d$. First, the axial ratio is decreased to its minimum Fig. 14(b) and next it increases with further increasing asymmetry Fig. 14(c). While the optimum axial ratio was observed in Fig. 14(b) at around $d = 500\ \mu m$, $Q$-balancing was found in Fig. 15 at around $d \approx 550\ \mu m$. Given the approximations in (35) (no dissipation loss and identical series and shunt patterns), this may be considered an excellent agreement. If the asymmetry is further increased, beyond the $Q$-balancing condition ($d > 550\ \mu m$), the shunt radiation power starts to dominate the series radiation power, following (29), which increases the off-broadside axial ratio.

## VI. CONCLUSION

Series and shunt radiation contributions in periodic LWAs have been defined in terms of equivalent voltages and shown to correspond to longitudinal and transverse polarization contributions, respectively. The respective contributions to polarization of the series and shunt radiation components have been investigated for axially symmetric and asymmetric periodic LWA structures. An intrinsic quadrature phase relationship between series and shunt radiation in transversally symmetric and longitudinally (axially) asymmetric periodic LWAs has been identified and theoretically proven, based on a pure symmetry argument. The case-study examples of a CRLH LWA and an SFP LWA have been presented to illustrate the quadrature phase relationship, corresponding to the first condition for circular polarization, has been established using circuit and full-wave simulation. The amplitude equality in the radiated fields, which represents the second condition, has been shown to be obtainable by controlling the degree of axial asymmetry.

The axial ratio of an SFP LWA has been minimized via a parametric study. Moreover, it has been empirically demonstrated that an optimum axial ratio corresponds to $Q$-balancing, so that an optimally circularly polarized LWA also features optimum radiation through broadside.

Little attention had been paid to the polarization characteristics and controllability in the literature. This work contributes to mend this deficiency and may stimulate the design of various types of LWAs with efficient circular polarization. The conclusions of the paper are not necessarily restricted to planar LWA configurations; they may also extend to waveguide and non-planar transmission-line LWAs.

## ACKNOWLEDGMENT

This work was supported by the "Ziel 2-Programm (EFRE) Transfer.NRW—Science-to-Business PreSeed."

## REFERENCES

[1] A. A. Oliner and D. R. Jackson, *Antenna Engineering Handbook*, J. Volakis, Ed., 4 ed. New York, NY, USA: McGraw-Hill, 2007, ch. 11.

[2] D. R. Jackson and A. A. Oliner, *Modern Antenna Handbook*, C. A. Balanis, Ed. New York, NY, USA: Wiley-Interscience, 2008.

[3] C. Caloz, D. R. Jackson, and T. Itoh, *Frontiers in Antennas*, F. Gross, Ed. New York, NY, USA: McGraw-Hill, 2010.

[4] C. Caloz and T. Itoh, *Electromagnetic Metamaterials: Transmission Line Theory and Microwave Applications*. Piscataway, NJ, USA: Wiley-IEEE Press, 2005.

[5] C. Caloz, T. Itoh, and A. Rennings, "CRLH metamaterial leaky-wave and resonant antennas," *IEEE Antennas Propag. Mag.*, vol. 50, no. 5, pp. 25–39, Oct. 2008.

[6] C. A. Allen, C. Caloz, and T. Itoh, "Leaky-waves in a metamaterial-based two-dimensional structure for a conical beam antenna application," in *IEEE MTT-S Int. Microwave Symp. Dig. (MTT)*, Jun. 2004, pp. 305–308.

[7] F. P. Casares-Miranda, C. Camacho-Penalosa, and C. Caloz, "High-gain active composite right/left-handed leaky-wave antenna," *IEEE Trans. Antennas Propag.*, vol. 54, no. 8, pp. 2292–2300, Aug. 2006.

[8] K. Mori and T. Itoh, "Distributed amplifier with CRLH-transmission line leaky wave antenna," in *Eur. Microwave Conf.*, Oct. 2008, pp. 686–689.

[9] K. Mori and T. Itoh, "CRLH metamaterial receiving leaky wave antenna integrated with distributed amplifier," presented at the Asia Pacific Microwave Conf., Hong Kong, Dec. 2008.

[10] T. Kodera and C. Caloz, "Uniform ferrite-loaded open waveguide structure with CRLH response and its application to a novel back-fire-to-endfire leaky-wave antenna," *IEEE Trans. Microwave Theory Tech.*, vol. 57, no. 4, pp. 784–795, Apr. 2009.

[11] C. M. Wu and T. Itoh, "A re-radiating CRLH transmission line leaky wave antenna using distributed amplifiers," presented at the Asia Pacific Microwave Conf., Singapore, Dec. 2009.

[12] S. Gupta, S. Abielmona, and C. Caloz, "Microwave analog real-time spectrum analyzer (RTSA) based on the spatial-spectral decomposition property of leaky-wave structures," *IEEE Trans. Microw. Theory Tech.*, vol. 59, no. 12, pp. 2989–2999, Dec. 2009.

[13] N. Yang, C. Caloz, and K. Wu, "Full-space scanning periodic phase-reversal leaky-wave antenna," *IEEE Trans. Microw. Theory Tech.*, vol. 58, no. 10, pp. 2619–2632, Oct. 2010.

[14] S. Abielmona, H. V. Nguyen, and C. Caloz, "Analog direction of arrival estimation using an electronically-scanned CRLH leaky-wave antenna," *IEEE Trans. Antennas Propag.*, vol. 59, no. 4, pp. 1408–1412, Apr. 2011.

[15] M. R. M. Hashemi and T. Itoh, "Evolution of composite right/left-handed leaky-wave antennas," *Proc. IEEE*, vol. 99, no. 10, pp. 1746–1754, Oct. 2011.

[16] T. Kodera, D. L. Sounas, and C. Caloz, "Non-reciprocal magnet-less CRLH leaky-wave antenna based on a ring metamaterial structure," *IEEE Antennas Wireless Propag. Lett.*, vol. 10, no. 1, pp. 1551–1554, Jan. 2012.

[17] D. R. Jackson, C. Caloz, and T. Itoh, "Leaky-wave antennas," *Proc. IEEE*, vol. 100, no. 7, pp. 2194–2206, Jul. 2012.

[18] W. J. Getsinger, "Elliptically polarized leaky-wave array," *IRE Trans. Antennas Propag.*, vol. 10, no. 2, pp. 165–171, Mar. 1962.

[19] K. Sakakibara, Y. Kimura, J. Hirokawa, M. Ando, and N. Goto, "A two-beam slotted leaky waveguide array for mobile reception of dual-polarization DBS," *IEEE Trans. Veh. Technol.*, vol. 48, no. 1, pp. 1–7, Jan. 1999.

[20] T. Hori and T. Itanami, "Circularly polarized linear array antenna using a dielectric image line," *IEEE Trans. Microw. Theory Tech.*, vol. 29, no. 9, pp. 967–970, Sep. 1981.

[21] C. T. Rodenbeck, M.-Y Li, and K. Chang, "Circular-polarized reconfigurable grating antenna for low-cost millimeter-wave beam-steering," *IEEE Trans. Antennas Propag.*, vol. 52, no. 10, pp. 2759–2764, Oct. 2004.

[22] Y. Dong and T. Itoh, "Realization of a composite right/left-handed leaky-wave antenna with circular polarization," in *Proc. Asia-Pacific Microwave Conf. Proc. (APMC)*, Dec. 2010, pp. 865–868.

[23] M. Hashemi and T. Itoh, "Circularly polarized composite right/left-handed leaky-wave antenna," in *Proc. IEEE Int. Conf. Wireless Inform. Technol. Syst. (ICWITS)* , Aug. 2010, pp. 1–4.

[24] N. Yang, H. Nguyen, S. Abielmona, and C. Caloz, "Mixed-mode characterization of a circularly-polarized CRLH leaky-wave antenna," in *Proc. 10th Int. Conf. Telecommun. in Modern Satellite Cable Broadcast. Services (TELSIKS)*, Oct. 2011, vol. 1, pp. 27–30.

[25] L. Geng, G.-M. Wang, H.-Y. Zeng, and M.-W. Chui, "Dual composite right/left-handed leaky-wave structure for dual-polarized antenna application," *Progress Electromagn. Res. Lett.*, vol. 35, pp. 191–199, 2012.

[26] H. V. Nguyen, S. Abielmona, and C. Caloz, "Highly efficient leaky-wave antenna array using a power-recycling series feeding network," *IEEE Antennas Wireless Propag. Lett.*, vol. 8, pp. 441–444, Mar. 2009.

[27] H. V. Nguyen, A. Parsa, and C. Caloz, "Power-recycling feedback system for maximization of leaky-wave antennas' radiation efficiency," *IEEE Trans. Microw. Theory Tech.*, vol. 58, no. 7, pp. 1641–1650, Jul. 2010.

[28] L. Liu, C. Caloz, and T. Itoh, "Dominant mode (DM) leaky-wave antenna with backfire-to-endfire scanning capability," *Electron. Lett.*, vol. 38, no. 23, pp. 1414–1416, Nov. 2002.

[29] S. Paulotto, P. Baccarelli, F. Frezza, and D. R. Jackson, "Full-wave modal dispersion analysis and broadside optimization for a class of microstrip CRLH leaky-wave antennas," *IEEE Trans. Microw. Theory Tech.*, vol. 56, no. 12, pp. 2826–2837, Dec. 2008.

[30] J. S. Gomez-Diaz, D. C. Rebenaque, and A. Alvarez-Melcon, "A simple CRLH LWA circuit condition for constant radiation rate," *IEEE Antennas Wireless Propag. Lett.*, vol. 10, pp. 29–32, Mar. 2011.

[31] S. Otto, A. Rennings, K. Solbach, and C. Caloz, "Transmission line modeling and asymptotic formulas for periodic leaky-wave antennas scanning through broadside," *IEEE Trans. Antennas Propag.*, vol. 59, no. 10, pp. 3695–3709, Oct. 2011.

[32] J. Gomez-Tornero, G. Goussetis, and A. Alvarez-Melcon, "Simple control of the polarisation in uniform hybrid waveguide-planar leaky-wave antennas," *IET Microwaves, Antennas Propag.*, vol. 1, no. 4, pp. 911–917, Aug. 2007.

[33] S Otto, A. Al-Bassam, A. Rennings, K. Solbach, and C. Caloz, "Radiation efficiency of longitudinally symmetric and asymmetric periodic leaky-wave antennas," *IEEE Antennas Wireless Propagation Lett.*, vol. 11, pp. 612–615, Jun. 2012.

[34] M. Ishii, T. Fukusako, and A. Alphones, "Design of leaky wave antenna with composite right-/left-handed transmission line structure for circular polarization radiation," *Progress Electromagn. Res. C*, vol. 33, pp. 109–121, 2012.

[35] J. D. Jackson, *Classical Electrodynamics*, 3rd ed. New York, NY, USA: Wiley, 1999.

[36] C. A. Balanis, *Antenna Theory: Analysis and Design*, 2nd ed. New York, NY, USA: Wiley, 1996.

[37] J. R. James and P. S. Hall, "Handbook of microstrip antennas," in *Ser. IEE Electromagnetic Waves Series, 28. INSPEC*, 2nd ed. London, U.K.: The Institution of Engineering and Technology, 1988.

[38] M. Zedler and G. Eleftheriades, "Spatial harmonics and homogenization of negative-refractive-index transmission-line structures," *IEEE Trans. Microw. Theory Tech.*, vol. 58, no. 6, pp. 1521–1531, Jun. 2010.

[39] C. Caloz and T. Itoh, "Array factor approach of leaky-wave antennas and application to 1-d/2-d composite right/left-handed (CRLH) structures," *IEEE Microw. Wireless Compon. Lett.*, vol. 14, no. 6, pp. 274–276, Jun. 2004.

[40] L. J. Chu, "Physical limitations of omni-directional antennas," *Appl. Phys.*, vol. 19, pp. 1163–1175, 1948.

[41] R. F. Harrington, *Time Harmonic Electromagnetic Fields*. New York, NY, USA: McGraw-Hill, 1961.

[42] R. E. Collin and S. Rothschild, "Evaluation of antenna Q," *IEEE Trans. Antennas Propag.*, vol. 12, no. 1, pp. 23–27, Jan. 1964.

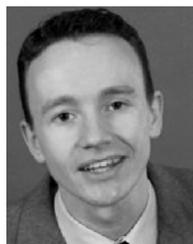

**Simon Otto** received the Diplom-Ingenieur degree in electrical engineering from Duisburg-Essen University, Essen, Germany, in 2004, where he is currently pursuing the Ph.D. degree.

Currently, he is with the Antenna and EM Modeling Department at IMST, Kamp-Lintfort, Germany. He was a visiting student at the Microwave Electronics Laboratory, University of California at Los Angeles (UCLA). He has authored or co-authored over 35 conference and journal papers related to antennas, filter designs, RF components, simulation techniques, magnetic-resonance imaging (MRI) systems, and filed five patents. His research interests include leaky wave antennas, periodic structures, transmission line metamaterials, EM-theory and numerical modeling.

Mr. Otto was a recipient of the VDE prize (Verband der Elektrotechnik Elektronik Informationstechnik e.V.) for his thesis and the second price of the Antenna and Propagation Symposium (AP-S) Student Paper Award 2005 in Washington.

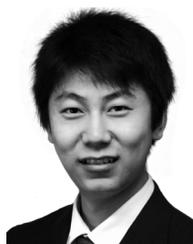

**Zhichao Chen** received the B.S. and M.S. degrees from the University of Duisburg-Essen, Essen, Germany, in 2009 and 2012, respectively. He is currently pursuing the Ph.D. degree in electrical engineering from the University of Duisburg-Essen, Germany.

His general research interests include array antennas, metamaterials, and electromagnetic theory. His current research focuses on development of RF components for magnetic-resonance imaging (MRI) systems.

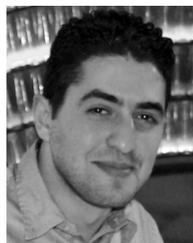

**Amar Al-Bassam** received the B.Sc. degree in electrical and electronic engineering from the University of Duisburg-Essen, Germany, in 2011, where he is currently pursuing the M.Sc. degree in electrical engineering.

Since October 2011, he has been working as a Student Assistant with the High-Frequency Technology (HFT) Department, University of Duisburg-Essen. In 2013, he was a visiting student at Poly Grames Research Center in Montréal, QC, Canada. His research interests include metamaterial-based leaky wave antennas and their applications.

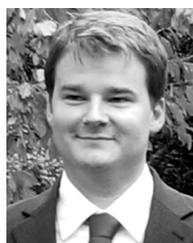

**Andreas Rennings** received his Dipl.-Ing. and Dr.-Ing. degrees from the University of Duisburg-Essen, Essen, Germany, in 2000 and 2008, respectively.

He studied electrical engineering with the University of Duisburg-Essen, Essen. He carried out his diploma work at the Microwave Electronics Laboratory, University of California at Los Angeles. From 2006 to 2008, he was with IMST GmbH, Kamp-Lintfort, Germany, where he worked as an RF Engineer. Since then, he is a Senior Scientist with the Laboratory for General and Theoretical Electrical Engineering, University of Duisburg-Essen, where he is a project leader in the field of bioelectromagnetics and med-tech. His general research interests include all aspects of theoretical and applied electromagnetics, currently with a focus on medical applications. He has authored and co-authored over 60 conference and journal papers and 1 book chapter and filed 8 patents.

Dr. Rennings was a recipient of several awards, including the second prize within the Student Paper Competition of the 2005 IEEE Antennas and Propagation Society International Symposium and the VDE-Promotionspreis 2009 for his doctoral thesis.

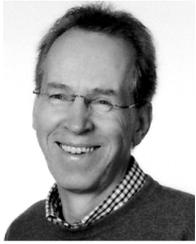

**Klaus Solbach** a junior researcher under Prof. Ingo Wolff in the field of integrated dielectric image line circuits with the University of Duisburg, Essen, Germany, from 1975 to 1980. In 1981, he joined the Millimeterwave Research Laboratory, AEG-Tele-funken, Ulm, Germany, and later changed to the Radar Systems Group, Daimler-Benz Aerospace (now part of EADS) in 1984, where he engaged in the design and production of microwave-subsystems for ground-based and airborne Radar, EW, and communication systems including phased array and active phased array antenna systems. His last position was Manager of the RF-and-Antenna-Subsystems Department. In 1997, he joined the faculty of the University of Duisburg as chair for RF and Microwave Technology. He has authored and co-authored more than 200 national and international papers, conference contributions, book chapters and patent applications.

Dr. Solbach was chairman of the VDE-ITG Fachausschuss "Antennen," executive secretary of the Institut fuer Mikrowellen-und Antennentechnik (IMA) and Chair of the IEEE Germany AP/MTT Joint Chapter. He was General Chair of the International ITG-Conference on Antennas INICA2007, Munich, Germany, and the General Chair of the European Conference on Antennas and Propagation EuCAP2009, Berlin, Germany.

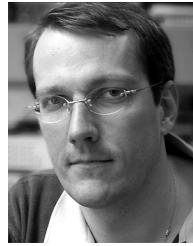

**Christophe Caloz** (F'10) received the Diplome d'Ingénieur en Électricité and the Ph.D. degree from École Polytechnique Fédérale de Lausanne (EPFL), Lausanne, Switzerland, in 1995 and 2000, respectively.

He is a Canada Research Chair. From 2001 to 2004, he was a Postdoctoral Research Engineer with the Microwave Electronics Laboratory, University of California at Los Angeles (UCLA). In June 2004, he joined the École Polytechnique of Montréal, Montréal, QC, Canada, where he is now a Full Professor, the holder of a Canada Research Chair (CRC) and the head of the Electromagnetics Research Group. He has authored and co-authored over 500 technical conference, letter, and journal papers, 12 books and book chapters, and he holds several patents. His works have generated more than 10 000 citations. In 2009, he co-founded the company ScisWave, which develops CRLH smart antenna solutions for WiFi. His research interests include all fields of theoretical, computational and technological electromagnetics, with strong emphasis on emergent and multidisciplinary topics, including particularly metamaterials, nanoelectromagnetics, exotic antenna systems and real-time radio.

Dr. Caloz is a Member of the Microwave Theory and Techniques Society (MTT-S) Technical Committees MTT-15 (Microwave Field Theory) and MTT-25 (RF Nanotechnology), a Speaker of the MTT-15 Speaker Bureau, the Chair of the Commission D (Electronics and Photonics) of the Canadian Union de Radio Science Internationale (URSI) and an MTT-S representative at the IEEE Nanotechnology Council (NTC). He was a recipient of several awards, including the UCLA Chancellor's Award for Post-doctoral Research in 2004, the MTT-S Outstanding Young Engineer Award in 2007, the E.W.R. Steacie Memorial Fellowship in 2013, the Prix Urgel-Archambault in 2013, and many Best Paper Awards with his students at international conferences.